\documentclass[prd,preprint,floatfix,nofootinbib]{revtex4}
\usepackage{epsfig}
\usepackage{graphicx}
\usepackage{pstricks}

\def\lsim{\mathrel {\vcenter {\baselineskip 0pt \kern 0pt
    \hbox{$<$} \kern 0pt \hbox{$\sim$} }}}
\def\gsim{\mathrel {\vcenter {\baselineskip 0pt \kern 0pt
    \hbox{$>$} \kern 0pt \hbox{$\sim$} }}}
\def\slashchar#1{\setbox0=\hbox{$#1$}           
 \dimen0=\wd0                                 
  \setbox1=\hbox{/} \dimen1=\wd1               
\ifdim\dimen0>\dimen1                        
  \rlap{\hbox to \dimen0{\hfil/\hfil}}      
  #1                                        
  \else                                        
 \rlap{\hbox to \dimen1{\hfil$#1$\hfil}}   
   /                                         
  \fi}                                         %
\def\cpto{\mathrel {\vcenter {\baselineskip 0pt \kern 0pt
    \hbox{$CP$} \kern 0pt \hbox{$\longrightarrow$} }}}
\def\cptof{\mathrel {\vcenter {\baselineskip 0pt \kern 0pt
    \hbox{$~CP$} \kern 0pt \hbox{$\longleftrightarrow$} }}}

\begin{document}

\baselineskip=15pt

\preprint{}

\title{An extended scalar sector to address the tension between a fourth generation and Higgs searches at the LHC}

\author{Xiao-Gang He$^{1,2}$ and German Valencia$^{3}$}

\email{hexg@phys.ntu.edu.tw,valencia@iastate.edu}

\affiliation{$^{1}$INPAC, Department of Physics, Shanghai Jiao Tong University, Shanghai, China\\
$^2$Department of Physics and Center for Theoretical Sciences, National Taiwan University, Taipei, Taiwan. \\
$^{3}$ Department of Physics, Iowa State University, Ames, IA 50011.}

\date{\today}

\vskip 1cm
\begin{abstract}

It is expected that the LHC will soon discover the Higgs boson, or that failure to find it will severely constrain its production cross-section over a large mass range. Either one of these results spells trouble for a fourth generation that significantly enhances the Higgs production cross-section at LHC.  In fact the LHC has already ruled out a SM Higgs mass in the range of 120 GeV to 600 GeV with a fourth generation at the 95\% C.L. In this paper we explore options within extended scalar sectors to maintain the viability of a heavy fourth generation if an enhanced (relative to the Standard Model) Higgs production cross-section is not observed.

\end{abstract}


\maketitle

\section{Introduction}

The search for the Higgs boson continues to be of paramount importance to complete our understanding of the standard model. In the last few years the CDF and D0 collaborations at the Tevatron have ruled out a mass window for the SM Higgs boson in the range $158-173$~GeV \cite{tevatron}.

The LHC has now joined the search for the Higgs boson, and relying on the $H\to WW$ mode, the ATLAS and CMS collaborations expect to either discover or exclude a standard model with three generations (SM3) Higgs boson in a wide mass range. Up to now, for SM3, a Higgs boson with a mass in the ranges $\sim 155 - 190$ and $\sim 295 - 450$~GeV  has been ruled out at 95\% C.L. by ATLAS \cite{atlas}, and in the ranges $\sim 149 - 206$ GeV and $\sim 300 - 400$ GeV by CMS \cite{cms}. With a fourth generation (SM4), CMS has ruled out the  Higgs boson mass in the range of \cite{cms} $\sim 120 - 600$ GeV at the 95\% C.L.

The dominant production mechanism for the SM3 Higgs boson at the LHC is gluon fusion \cite{Spira:1995rr} through a top-quark loop. This mechanism is very sensitive to new physics: for example, with a fourth generation (SM4), the amplitude is roughly 3 times larger as it simply counts the number of heavy flavors in the loop. This results in a significantly larger cross-section for Higgs boson production, about 9 times larger than in SM3 \cite{Li:2010fu}.

This observation, combined with the Higgs search results from the LHC this year and expected from future, places a considerable strain on the possibility of a fourth generation. If the Higgs boson is indeed discovered with a production cross-section roughly in agreement with the SM, one would have to explain why the factor of 9 enhancement present in SM4 is not there. On the other hand, if the Higgs boson is not observed, its exclusion is even more significant in SM4, and its allowed mass is pushed towards the unitarity bound.
A possible way out of this predicament that has been discussed for SM4 is that it may be more natural to have a heavy Higgs if indeed there are four generations \cite{Holdom:2010za}.

In this note we consider the possibility of extended scalar sectors that could remove the tension between Higgs physics at the LHC and a heavy fourth generation. We discuss two possibilities:  a scalar sector extended with a color octet, electroweak doublet \cite{Manohar:2006ga}; and a variation of the two-Higgs doublet model ``for the top'' \cite{Das:1995df,BarShalom:2011zj}. In the first case we argue that it is possible to suppress the Higgs production rate in gluon fusion with suitable additional particles. In the second case we argue that it is possible to make the neutral scalar with  SM-Higgs-like couplings heavy, or to suppress its production cross-section relative to SM4. Similar arguments have been used in Ref.~\cite{Gunion:2011ww} to constrain a possible fourth generation in two-Higgs doublet models of type II. Another possibility recently discussed occurs in models where the Higgs boson has a larger invisible decay width than that in the SM~\cite{invisible}.

\section{Color octet scalars}

We now consider the case of a scalar sector that has been extended with a color octet, electroweak doublet and hypercharge $1/2$ scalar $O=(8,2,1/2)$. This particular choice is
motivated by the requirement of minimal flavor violation and has been recently elaborated in Ref.~\cite{Manohar:2006ga}. It
was noted in that paper that in minimal flavor violation only scalars with the same gauge quantum numbers as the SM Higgs doublet $H = (1,2,1/2)$
or color octet scalars with the same weak quantum numbers as the Higgs doublet $O = (8,2,1/2)$ can couple to quarks, a scenario  with  many interesting collider and flavor physics consequences. This color octet can be written in a conventionally normalized component form with color index $A$ as $O = \sqrt{2} S =\sqrt{2} T^A (S^{A+}, S^{A0})^T$.

Since the new scalars carry color, they may have a significant effect on the two gluon coupling to the SM Higgs boson $h$, and thus affect its production at LHC.
The contribution of $S$ to the $h-gg$ coupling cannot happen at the tree level, but occurs at the one loop level by having the color octet $S$ in the loop and allowing $h$ to couple to $S$ from the interactions in the scalar potential.
The most general scalar potential with $H$ and $S$ is Ref.\cite{Manohar:2006ga},
\begin{eqnarray}
V &=& {\lambda\over 4} \left ( H^{\dagger i} H_i - {v^2\over 2}\right )^2 + 2 m^2_S {\rm Tr}S^{\dagger i}S_i + \lambda_1 H^{\dagger i} H_i {\rm Tr}S^{\dagger j}S_j + \lambda_2 H^{\dagger i}H_j
{\rm Tr} S^{\dagger j}S_i \nonumber\\
&+& \left[ \lambda_3 H^{\dagger i} H^{\dagger j} {\rm Tr}S_i S_j + \lambda_4 H^{\dagger i}{\rm Tr}S^{\dagger j}S_j S_i +  \lambda_5 H^{\dagger i}{\rm Tr}S^{\dagger j}S_i S_j + {\rm ~H.c} \right]\nonumber\\
&+&\lambda_6 {\rm Tr} S^{\dagger i} S_i S^{\dagger j}S_j + \lambda_7 {\rm Tr} S^{\dagger i} S_j S^{\dagger j}S_i + \lambda_8 {\rm Tr} S^{\dagger i} S_i{\rm Tr} S^{\dagger j}S_j \nonumber\\
&+&\lambda_9 {\rm Tr} S^{\dagger i} S_j {\rm Tr} S^{\dagger j}S_i + \lambda_{10} {\rm Tr} S_i S_j S^{\dagger i}S^{\dagger j} + \lambda_{11} {\rm Tr} S_i S_j S^{\dagger j}S^{\dagger i}\;.
\end{eqnarray}

The additional contribution to the Higgs boson production due to the octet-scalar loops has been calculated in  Ref.~\cite{Manohar:2006ga}. In the limit of very heavy quarks and color scalars in the loop, the  $h-gg$ coupling can be obtained form the effective Lagrangian\begin{eqnarray}
{\cal L} = (\sqrt{2} G_F)^{1/2} \frac{\alpha_s}{12\pi}\
G^A_{\mu\nu}G^{A\mu\nu} h \left(n_{hf} +\frac{v^2}{m^2_S}\frac{3}{8}(2\lambda_1+\lambda_2)\right)
\label{effhgg}
\end{eqnarray}
where $v$ is the vacuum expectation value (vev) of the usual SM Higgs doublet $H = (1,2,1/2) = (H^+, (v+h+i I)/\sqrt{2})$. $H^+$ and $I$ are the would-be Goldstone bosons ``eaten'' by $W^+$ and $Z$, and $h$ is the physics Higgs field. $n_{hf}$ is the number of heavy quark flavors, one in the case of SM3 and three in the case of SM4.

It is clear from the above expression for the $gg-h$ coupling that the addition of the color octet scalar doublet has the potential to significantly alter the Higgs production cross-section at LHC. Indeed, it was already pointed out in Ref.~\cite{Manohar:2006ga} that this cross section could double for $\lambda_1=4,\ \lambda_2=1$, which are consistent with electroweak constraints. It is equally possible to cancel an enhancement from SM4 in the Higgs production cross-section, $n_{hf}=3$, with appropriate parameters in the color octet scalar potential.   For this cancellation to be possible, we need to check that the required values of $\lambda_{1,2}$ are not in conflict with anything else.

There is no constraint from the electroweak parameter $T$ (equivalently from custodial $SU(2)$) on the  combination $(2\lambda_1+\lambda_2)$, as long as a third parameter in the potential, $\lambda_3$, takes the value $\lambda_3 = 2\lambda_2$. The contribution from $\lambda_2$ to the electroweak parameter $S$ is given by \cite{Manohar:2006ga}
\begin{eqnarray}
\lambda_2 &=& 6\pi \frac{m^2_S}{v^2} S.
\end{eqnarray}

The current best fit value for S parameter and its allowed range, for $M_h=300$~GeV is,
$S = -0.07 \pm 0.09$ \cite{Nakamura:2010zzi}.
Assuming that $S$ is saturated by $\lambda_2$, the implied range for the combination appearing in Eq.~\ref{effhgg} is
\begin{equation}
\frac{v^2}{m_S^2}\frac{3}{8}\lambda_2 = -0.49 \pm 0.64
\end{equation}
This is enough for a large suppression of SM3 at one sigma level. To sufficiently suppress SM4, an additional negative contribution from $\lambda_1$,
which is  not constrained by $S$, is required. For example, with $m_S\sim 2 v$, a $\lambda_1\sim -8$ would halve the SM4 coupling.
Of course, the color octet scalar can also enhance the Higgs production cross section, exacerbating any potential conflict.

\section{A two Higgs doublet model}

Another simple extension of the scalar sector of the SM  consists of adding a second Higgs doublet \cite{Donoghue:1978cj}. Many variations on this theme exist in the literature, but one stands out motivated by the fact that the fourth generation quarks are necessarily much heavier than the lightest five quarks. This suggests that a second Higgs doublet is responsible for the masses of the fourth generation quarks (and possibly the top-quark as well), in a variation of the ``for the top-quark'' two Higgs doublet model of Ref.~\cite{Das:1995df,BarShalom:2011zj}. The phenomenology of these models and their significance for a heavy fourth generation has been recently emphasized in Ref.~\cite{BarShalom:2011zj}.

To implement this model as a solution to the SM4 tension with Higgs physics at the LHC, we do not focus on reducing the $gg-h$ effective  coupling. We consider  instead the observability of $h$ including its production and its couplings to  $W$ pairs as well as its mass. In essence, we require the Higgs with relatively large couplings to $W$-pairs either to be outside the accessible mass range, or to have a smaller enhancement over SM3 than what is found in SM4.

In a generic two Higgs doublet model with scalar fields $H_1$ and $H_2$ each has a vev $v_1$, $v_2$.
Assuming that $H_1$ couples to the first three generations and $H_2$ couples to the fourth generation, the Yukawa couplings are given by (we call this Model I)
\begin{eqnarray}
{\cal L} &=& - \bar Q^i_L Y^u_{ij} \tilde H_1 U^j_R -\bar Q^i_L Y^d_{ij}  H_1 D^j_R -\bar L^i_L Y^e_{ij} H_1 E^j_R +{\rm ~H.c.} \nonumber\\
&&- \bar Q^4_L Y^u_{44} \tilde H_2 U^4_R -\bar Q^4_L Y^d_{44}  H_2 D^4_R -\bar L^4_L Y^e_{44} H_2 E^4_R + {\rm ~H.c.}
\end{eqnarray}
The model can be modified so that  the top-quark also couples to $H_2$, we call this Model II.

After spontaneous symmetry breaking there remain two physical neutral scalars $h$ and $H$,  a pseudo-scalar $A$, and a charged Higgs $H^+$. The physical neutral scalars are the ones relevant to our discussion here. In general, the Higgs potential parameters  mix the neutral real components of the doublets $h^0_{1,2}$ to form the physical neutral scalars $h$ and $H$. We parameterize this mixing by
\begin{eqnarray}
\left ( \begin{array}{l} h\\ H
\end{array}\right ) = \left ( \begin{array}{ll} \cos\alpha&\sin\alpha\\ -\sin\alpha&\cos\alpha\end{array}\right ) \left ( \begin{array}{l} h_1\\ h_2
\end{array}\right )\;.
\end{eqnarray}
At this point it is worth noting that the parameter space in the Higgs potential can accommodate either  $h$ or $H$ as the heavier of the two.

The couplings of $h$ and $H$ to $W$ and $Z$ bosons are given by
\begin{eqnarray}
{\cal L} = \left(2{m^2_W\over v} W^{+\mu}W^-_\mu + {m^2_W\over v}Z^\mu Z_\mu\right) \left( H\sin(\beta - \alpha) + h\cos(\beta - \alpha)\right)\;,
\end{eqnarray}
where $\beta$ is defined by $\tan^{-1}(v_2/v_1)$ and $v^2 = v^2_1+v^2_2$, and $\tan\beta$ is presumably large as $v_2$ gives mass to the heavy fourth generation.

The Yukawa couplings of $h$ and $H$ are given by ($i=1,2,3$)
\begin{eqnarray}
{\cal L} &=& - {1\over v} (\bar u^i m^u_i u^i + \bar d^i m^d_i d^i + \bar e^i m^e_i e^i)\left({\cos\alpha\over \cos\beta} h - {\sin\alpha\over \cos\beta}H\right)\nonumber\\
&&- {1\over v} (\bar u^4 m^u_4 u^4 + \bar d^4 m^d_4 d^4 + \bar e^4 m^e_4 e^4)\left({\sin\alpha\over \sin\beta} h + {\cos\alpha\over \sin\beta}H\right)\;.
\end{eqnarray}

The LHC search strategy  for the Higgs masses in the (120-600)~GeV range mentioned before, corresponds to the process $gg \to H$ followed by $H \to W^+W^-$ or $H \to ZZ$ \cite{cms}. With the couplings of Model I described above, the cross-sections for the overall processes, as compared to the SM3 Higgs, become (in the infinite heavy quark mass limit)\footnote{Notice that the expressions in Eq.~\ref{ratios} are the same in the 2HDM type II, although the origin of the separate terms is different: the term with the factor of 2 arising from two up-type quarks $t,t^\prime$ and the second term arising from the $b^\prime$ quark. }
\begin{eqnarray}
\frac{\sigma_h}{\sigma_{SM3}} \equiv \frac{\sigma(pp\to h \to VV)}{\sigma(pp\to H_{SM} \to VV)} &\sim &
\left[\left(2\frac{\sin\alpha}{\sin\beta}+\frac{\cos\alpha}{\cos\beta}\right)\cos(\beta-\alpha)\right]^2 \nonumber \\
\frac{\sigma_H}{\sigma_{SM3}} \equiv \frac{\sigma(pp\to H \to VV)}{\sigma(pp\to H_{SM} \to VV)} &\sim &
\left[\left(2\frac{\cos\alpha}{\sin\beta}-\frac{\sin\alpha}{\cos\beta}\right)\sin(\beta-\alpha)\right]^2
\label{ratios}
\end{eqnarray}
where $V=W {\rm ~or~} Z$. If we choose to couple the top-quark to $H_2$ instead of $H_1$ as in Model II, these ratios become
\begin{eqnarray}
\frac{\sigma_h}{\sigma_{SM3}} \equiv \frac{\sigma(pp\to h \to VV)}{\sigma(pp\to H_{SM} \to VV)} &\sim &
\left[\left(3\frac{\sin\alpha}{\sin\beta}\right)\cos(\beta-\alpha)\right]^2 \nonumber \\
\frac{\sigma_H}{\sigma_{SM3}} \equiv \frac{\sigma(pp\to H \to VV)}{\sigma(pp\to H_{SM} \to VV)} &\sim &
\left[\left(3\frac{\cos\alpha}{\sin\beta}\right)\sin(\beta-\alpha)\right]^2.
\label{ratios2}
\end{eqnarray}

\begin{figure}[th]
\centerline{
\includegraphics[width=.4\textwidth]{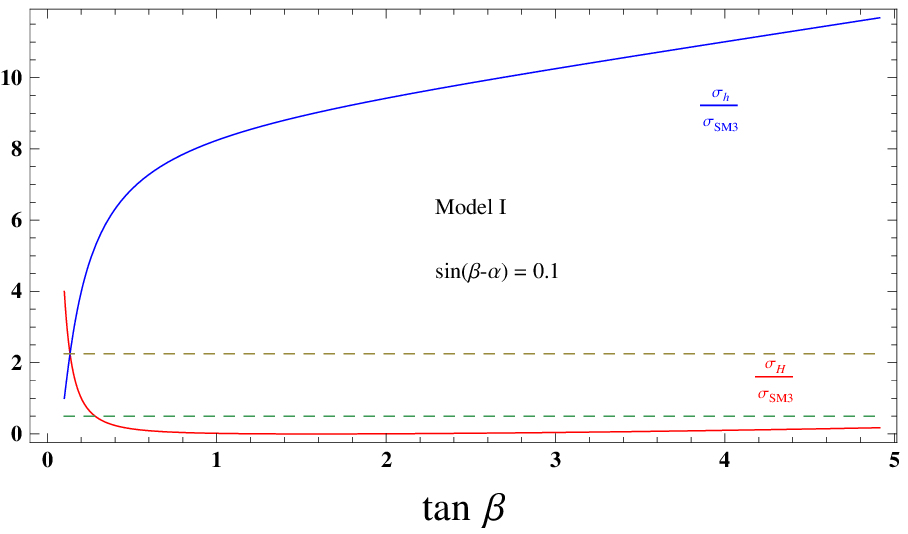}
\includegraphics[width=.4\textwidth]{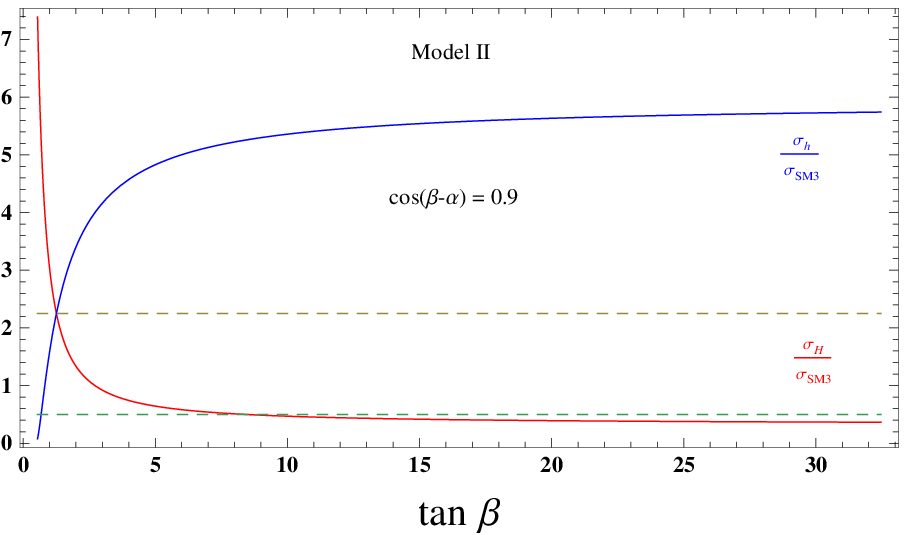}
} \caption{Ratios for Model I in Eq.~\ref{ratios} (left plot) and for Model II in  Eq.~\ref{ratios2} (right plot) as a function of  $\tan\beta$. The dashed lines mark the points $2.25$, where the two ratios are equal; and $0.5$, where the cross-section is half the SM3 value. }\label{fig:ratios}
\end{figure}

In Figure~\ref{fig:ratios} we plot the ratios in Eq.~\ref{ratios} (Model I) and Eq.~\ref{ratios2} (Model II) for a region of parameter space. We use $\sin(\beta-\alpha)=0.1$ for Model I and $\cos(\beta-\alpha)=0.9$ for Model II which make the $h$ coupling to $W$ pairs very similar to the SM Higgs coupling to $W$ pairs. We plot ranges for $\tan\beta$ which are similar to those considered in Ref.~\cite{BarShalom:2011zj}. A full analysis of the parameter space of these models is beyond the scope of this note, but our figure illustrates the salient features that can alleviate the tension with SM4.

One possibility is to make the SM-like Higgs boson heavy. The main difference between this scenario and a heavy Higgs in SM4 is that the second, lighter, Higgs increases the mass bound allowed by unitarity for the heavy one \cite{unitarity}. For example, consider a case in which the $h$ is the SM-like Higgs boson such that its coupling to $W$-pairs is close to that in the SM.  In this limit, the $h$ couplings to the fermions are also the same as those for the SM4 and $\sigma_h/\sigma_{SM3}\sim 9$. It follows that if  $h$ is also the lightest neutral scalar eigenstate, this model is identical to and has the same tension as SM4. However, in this 2HDM, the mass of $h$ can be higher than the mass of the  Higgs boson of the SM4. In particular, generic unitarity bounds for two-Higgs doublet models suggest it can be as heavy as $\sim 700$~GeV \cite{unitarity} and outside the range of current searches.  In this scenario, $H$ is the lighter Higgs boson and can be produced at the LHC, but the search for this object requires a different strategy as it does not couple to $W$-pairs, $\sigma_H/\sigma_{SM} << 1$. In Figure~\ref{fig:ratios} we illustrate this case for Model I, Eq.~\ref{ratios}, on the left plot. The figure shows that there is a value of $\tan\beta$ where  $\sigma(pp\to h,H \to VV)$ is the same for both Higgs bosons, and only 2.25 times larger than it is in SM3.  As we increase $\tan\beta$ from that point, the ratio $\sigma_h/\sigma_{SM3}$ increases until it reaches the value of 9 as in SM4. At the same time the ratio $\sigma_H/\sigma_{SM3}$ is decreasing being less than half the SM3 value for most of the range.  This illustrates a second way to alleviate the tension, in which the SM-like Higgs boson is still enhanced relative to SM3 but by a smaller factor than it is in SM4.

In the right plot of Figure~\ref{fig:ratios} we show what happens in Model II (where the top-quark also couples to $H_2$) for $\cos(\beta-\alpha)=0.9$. Notice that it is not possible to find values of $\alpha$ and $\beta$ that simultaneously suppress $\sigma_H/\sigma_{SM3}$ and  $\sigma_h/\sigma_{SM3}$ in Eq.~\ref{ratios2} below one.   However, the enhancement over SM3 can be made as small as 2.25 simultaneously for both neutral Higgs bosons. The figure also shows that one can suppress the lighter neutral Higgs below the SM3 value while enhancing the heavier one by factors around 5, which is still below SM4. 

In summary, in a generic two-Higgs doublet model, the process $gg\to h (H) \to W W$ will not be able to conclusively rule out a heavy fourth generation unless the search is extended to masses reaching the unitarity bound and the sensitivity to the cross-section reaches values that are close to twice the SM3 cross-section.

\section{Conclusion}

The search for the Higgs boson at the LHC during the upcoming year will be at odds with the predictions of a heavy fourth generation if the Higgs boson is found with a cross-section consistent with SM3 or if the Higgs boson is not found at all.
In this note we have examined the possibility of ameliorating this conflict with extended scalar sectors. There are two possible paths that we illustrate with two examples. In the first example, an additional color-octet electroweak-doublet can suppress the Higgs production  cross-section in both SM3 and SM4. The details, of course, depend on the parameters in the scalar potential but we have seen that the numbers needed are allowed by other constraints. In the second example we have shown that in a two-Higgs doublet model there is sufficient freedom to suppress the Higgs cross-section for the lighter neutral Higgs, while at the same time the unitarity bounds allow the mass of the heavier Higgs boson  to be larger than in SM4, possibly outside the current LHC search range.

\begin{acknowledgments}

This work was supported in part by DOE under contract number DE-FG02-01ER41155.
G.V. thanks the Physics Department at the National Taiwan University, Taipei, Taiwan for their hospitality
while this work was completed. XGH was partially supported by NSC and NCTS od ROC, and SJTU 985 grant of PRC.

\end{acknowledgments}

\appendix

\end{document}